\documentclass{aa}

\usepackage{graphicx}
\usepackage{txfonts}

\usepackage{braket}
\usepackage{amsmath}
\usepackage{bm}
\usepackage[colorlinks,allcolors=blue]{hyperref}
\usepackage[a]{esvect}
\makeatletter
\renewcommand*\aa@pageof{, page \thepage{} of \pageref*{LastPage}}

\makeatother
\usepackage{amstext}
\usepackage[normalem]{ulem}

\begin{document}

\title{Using deep learning to characterize single-exposure double-line spectroscopic binaries}

\author{ 
A. Binnenfeld\inst{\ref{inst2}}
\and S. Lilek\inst{\ref{inst1}} 
\and R. Nasser\inst{\ref{inst9}}
\and R.  Giryes\inst{\ref{inst1}}
\and S. Zucker\inst{\ref{inst2}, \ref{inst3}}}

\institute{
Porter School of the Environment and Earth Sciences, Raymond and Beverly Sackler Faculty of Exact Sciences, 
Tel Aviv University, Tel Aviv, 6997801, Israel \\
\email{avrahambinn@gmail.com,shayz@tauex.tau.ac.il}\label{inst2}
\and
School of Electrical and Computer Engineering, Iby and Aladar Fleischman Faculty of Engineering, Tel Aviv University, Tel Aviv, 6997801, Israel \label{inst1}
\and
School of Computer Science and AI, Tel Aviv University, Tel Aviv, 6997801, Israel 
\label{inst9}
\and
School of Physics and Astronomy, Raymond and Beverly Sackler Faculty of Exact Sciences, Tel Aviv University, Tel Aviv, 6997801, Israel\label{inst3}}

% These dates will be filled out by the publisher
% \date{Accepted XXX. Received YYY}

\abstract{
Distinguishing the component spectra of double-line spectroscopic binaries (SB2s) and extracting their stellar parameters is a complex and computationally intensive task that usually requires observations spanning several epochs that represent various orbital phases. This poses an especially significant challenge for large surveys such as \textit{Gaia} or LAMOST, where the number of available spectra per target is often not enough for a proper spectral disentangling. We present a new approach for characterizing SB2 components from single-exposure spectroscopic observations. The proposed tool uses deep neural networks to extract the stellar parameters of the individual component spectra that comprise the single exposure, without explicitly disentangling them or extracting their radial velocities. The neural networks were trained, tested, and validated using simulated data resembling \textit{Gaia} RVS spectra, which will be made available to the community in the coming \textit{Gaia} data releases. We expect our tool to be useful in their analysis. }


\titlerunning{Using deep learning to characterize single-exposure double-line spectroscopic binaries}
\authorrunning{A. Binnenfeld}

\maketitle

\section{Introduction}
\label{sec:intro}

The analysis of double-line spectroscopic binaries (SB2s) typically involves an initial characterization of the spectra of individual orbiting components and an estimation of their respective stellar parameters. These tasks are needed to study the physical properties of SB2 components \citep[e.g.][]{Kounkel_2021} and to extract and analyze their radial velocities (RVs) \citep{hill1993, zucker94}.

Several methods were used in the past to separate the component spectra of SB2s. \cite{bangies1991} introduced a tomographic approach to the problem. The approach treats the varying RVs of the components as different viewing angles of a blended spectral object. By analyzing spectra at multiple orbital phases, this method disentangles the composite spectra and reveals the individual spectral features of each star, which is much like reconstructing a three-dimensional object from its images that were obtained from different perspectives. The method makes use of the iterative least-squares technique (ILST) introduced by \cite{Brooks_1976}.

One of the most commonly used disentangling methods was introduced by \cite{Simon1994}. Assuming a known constant luminosity ratio, the authors approached disentangling as a linear problem. This allowed the problem to be solved using a least-squares-based standard singular-value decomposition (SVD) method. The strength of this approach lies in the fact that it does not require assumptions on any spectral model or a prior extraction of RVs.

\cite{KOREL} introduced the KOREL code for the simultaneous decomposition of component spectra and a solution of orbital elements. In this method, the observations are fit to superpositions of spectra that are attributed to individual components, shifted according to different RV values. Unlike the approach by \cite{Simon1994}, which is based on the decomposition of spectra in the wavelength domain, KOREL uses the least-squares fit of Fourier transforms of the observed spectra. This makes the solution more efficient numerically.

Recently, \citet{2018MNRAS.473.5043E} developed a method enabling the identification of binary stars and the extraction of their component parameters in an Apache Point Observatory Galactic Evolution Experiment (APOGEE) spectrum by fitting a synthetic binary model. This method was subsequently applied to a large sample from APOGEE DR13 \citep{2018MNRAS.476..528E}. Similar tools following this approach were later introduced for Gaia-European Southern Observatory  \citep[Gaia-ESO;][]{2012Msngr.147...25G, 2013Msngr.154...47R} by \citet{2022MNRAS.510.1515K}, and for Large Sky Area Multi-Object Fiber Spectroscopic Telescope (LAMOST) Medium-Resolution Spectroscopic Survey \citep[LAMOST-MRS;][]{liu2020lamost} by \citet{2022MNRAS.517..356K, 2024MNRAS.527..521K}.

Deep-learning-based machine-learning methods have frequently been used for a wide range of astronomical applications \citep[see][]{baron19}. Examples include the detection of periodic exoplanet transits in the presence of colored noise \citep{zuck2018,dvash2022}, exoplanet discovery in direct imaging with a low signal-to-noise ratio (S/N) \citep{Yip2020}, and the detection of objects of interest in spectroscopic surveys \citep{Skoda2020}.

Several deep-learning methods have been proposed for determining stellar parameters from stellar spectra. This task was previously addressed using a variety of techniques \citep{Kassounian19}. A comprehensive review of more than two decades of research in this field is provided by \citet{Gebran22}. These efforts include diverse neural network architectures that were applied to stellar spectral classification and parametrization for several decades \citep[e.g.,][]{vonHip, Bailer_Jones_1997}. The dramatic improvements in neural network performance since the release of AlexNet \citep{alex_net} have positioned deep learning as a powerful tool for addressing new challenges in stellar spectroscopy.

In the context of an SB2 analysis, \cite{zhang2021spectroscopic} used a convolutional neural network (CNN) to detect SB2s based on a single-exposure medium-resolution spectrum. The method presented by \citeauthor {zhang2021spectroscopic} did not characterize the detected SB2s further than tagging them as SB2s for further research. No attempt was made to measure the RVs or disentangle the component spectra. The method was tailored for the LAMOST-MRS. Subsequently, \citet{2024arXiv241103994J} applied the method to LAMOST Low-Resolution Spectra (LAMOST-LRS) as well.

\textit{Gaia} is a current space mission by the European Space Agency (ESA) that was launched in 2013. It has provided high-precision positions, parallaxes, and motions for more than $1.8$ billion stars so far \citep{gaia2016, gaiadr3}. Although it mainly is an astrometric mission, \textit{Gaia} is also equipped with other instruments, including the Radial Velocity Spectrograph (RVS), which is obtaining spectra for relatively bright stars included in the survey ($G < 15.5$), with a resolution of $R = 11,000$ \citep{2018Cropper}. 

This paper presents a study to determine whether deep learning can extract the stellar parameters of SB2 components based on a single-epoch \textit{Gaia} RVS spectral observation, without explicitly disentangling their spectra or extracting their RVs. Using synthetic data, we trained and validated a designated network capable of performing this task. 

The structure of the paper is as follows: Sect.~\ref{sec:sim_data} presents the way we generated the synthetic data. Sect.~\ref{sec:training} details the development and architecture of the neural network itself, and Sect.~\ref{sec:RES} presents the results we obtained using our model. We summarize our findings in Sect.~\ref{sec:diss} and discuss the method and its potential for future study.

\section{Data simulation}
\label{sec:sim_data}
\textit{Gaia} RVS is obtaining spectra at a resolution of $R = 11,000$ in the spectral range $8470-8710$\,\AA. We generated data of simulated SB2 spectra in this range using the PHOENIX spectral library \citep{Husetal2013}.

The stellar parameter components, including the effective temperature, $\log g$, and metallicity ($[\element{Fe}/\element{H}]$), were randomly drawn from the ranges presented in Table \ref{table:comp}. Although $\alpha$-element abundance is also available in the PHOENIX spectral library, it was excluded from this work because it was not computed for all effective temperature and metallicity ranges in the library \citep[see][]{Husetal2013}. 

The spectral parameters can assume a continuous range of values, but the PHOENIX data are provided as a discrete grid of these quantities. We therefore interpolated the spectra using different weights to emulate a continuous variation. In order to avoid introducing spectral artifacts, we interpolated each spectrum along only one axis of the PHOENIX grid at a time (either $T_\text{eff}$, $\log g$, or $[\element{Fe}/\element{H}]$) while keeping the other parameters fixed at their original grid values. This approach allowed us to generate intermediate values for a given parameter while preserving the physical consistency of the synthetic spectra. We further broadened the individual spectra assuming a rotational velocity of $v \sin i= 0 - 100\,\mathrm{km\,s}^{-1}$ using the code package PyAstronomy \citep{pya}. 

Similarly to the simulation presented by \citet{binnenfeld20}, we shifted and blended the spectra, assuming circular orbits with different periods. The orbital orientation was determined so that the maximum RV separation ($K_1 + K_2$) would range between $ 10 - 90\,\mathrm{km\,s}^{-1}$. In order to determine the individual semi-amplitudes $K_1$ and $K_2$, as well as the intensity ratio for combining the spectra, we used the masses and radii listed in PHOENIX, assuming main-sequence stellar properties. We sampled each simulated orbit at a single epoch, and we added noise with an S/N in the range of $30 - 350$. 

\begin{table}
\caption{Range of stellar parameters in the training data.}
\label{table:comp}
\begin{tabular}{lllccc}
\hline
\multicolumn{1}{l}{ } & 
\multicolumn{1}{l}{$T_\text{eff} [\mathrm{K}]$} & 
\multicolumn{1}{c}{$\log g$} & 
\multicolumn{1}{c}{$[\element{Fe}/\element{H}]\, [\text{dex}]$}  &
             % \multicolumn{1}{c}{\text{Alpha}} &
\multicolumn{1}{c}{$v \sin i \, [\text{km}\,\text{\ s}^{-1}]$} \\
\hline
\noalign{\smallskip}
\textrm{Lower} & $3500$ & $3.0$ & $-4.0$ & $0$ \\
\textrm{Upper} & $6500$ & $6.0$ & $+1.0$ & $100$ \\
\noalign{\smallskip}
\hline
\noalign{\smallskip}
\end{tabular}
\end{table}

\begin{figure} % [width=\columnwidth,clip=true]
\centering
\includegraphics[width=0.85\columnwidth,clip=true]{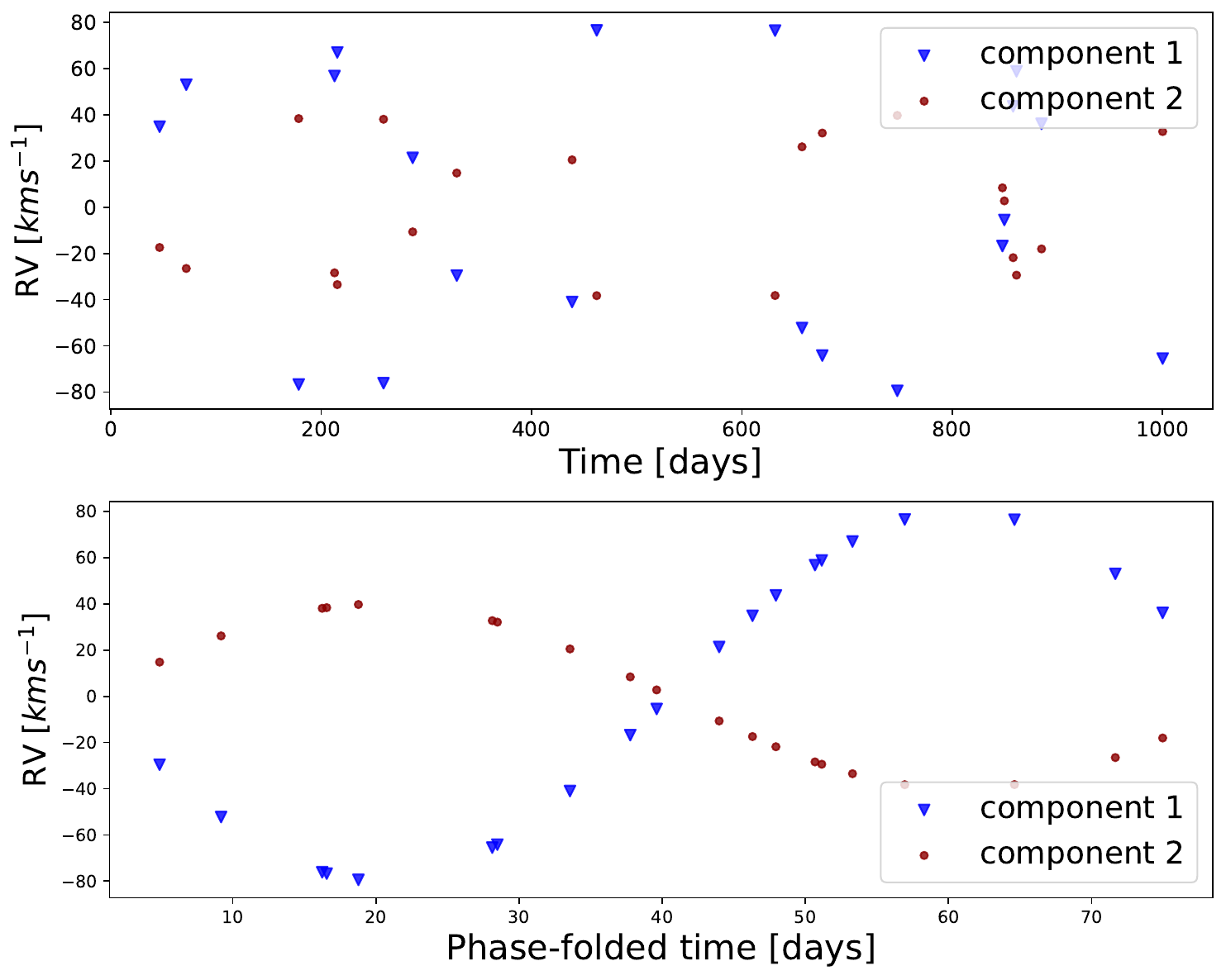}
\caption{Example of a characteristic 20-epoch RV time series of an SB2 system generated in the process described in Sect.\,\ref{sec:sim_data}. The red circles mark the primary RV, and the blue triangles show the RV of the secondary. Top panel: RVs plotted against observation times. Bottom panel: Same RV time-series, phase-folded by the known 81-day period.}
\label{fig:comps_example_rvs}
\end{figure}

\begin{figure*}
\centering
\includegraphics[width=1.9\columnwidth,clip=true]{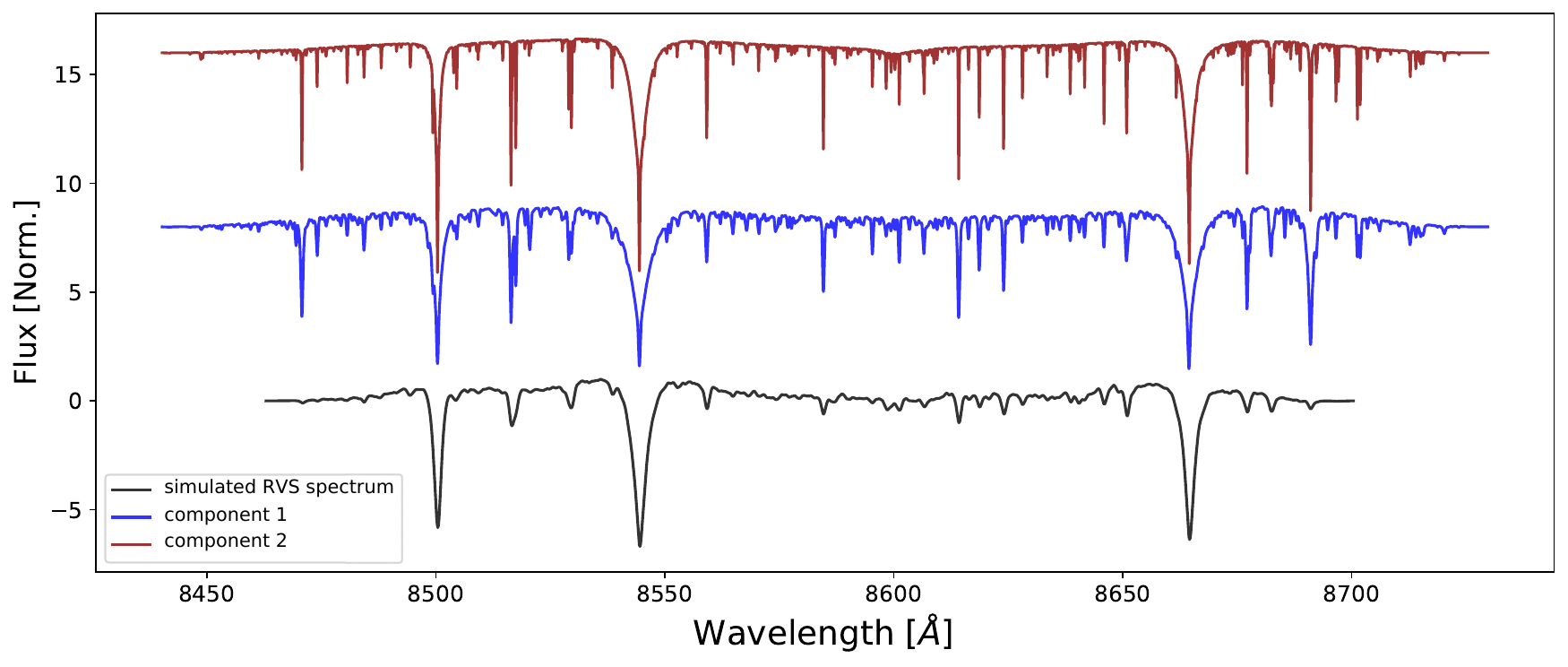}
\caption{Underlying spectra of the two components from the simulated system presented in Fig. \ref{fig:comps_example_rvs} at a random epoch (red and blue upper spectra), next to the resulting observed spectrum (lower spectrum in black).
The spectral parameters for the two components are a $T_\text{eff}$ of $5,500$, and $6,500$ $ \mathrm{K}$, a $\log g$ of 6.0 and 5.0, $\element{Fe}/\element{H}\, $ of $ 0.5$ and $ 0$ $ \text{dex}$, and a $v \sin i \, $ of $ 9 $ and $ 2$ $ \text{km}\,\text{\ s}^{-1}$, respectively. %\\
The spectra are normalized to a continuum level of zero. To enhance clarity, the spectra are presented here in arbitrary units, and a vertical offset was applied to the plot to separate them visually. }
\label{fig:comps_example}
\end{figure*}

Fig.~\ref{fig:comps_example_rvs} presents an example of a characteristic $20$-epoch RV time series of an SB2 system generated in the above process. Fig.~\ref{fig:comps_example} presents the underlying spectra of both components at a random epoch, next to the resulting combined spectrum. Note that missing features near the spectrum edges, due to processing, have little impact on similarly simulated data but should be considered for real data, for instance, by extending the training wavelength range. We generated $200$,$000$ RVS epoch spectra of SB2 systems generated in the above way, and we divided them into training, test, and validation datasets (see Sect. \ref{sub-training_det}).

\section{Model development}
\label{sec:training}

To extract the stellar parameters of the SB2 components, we employed the DenseNet \citep{huang2018densely} Convolutional Neural Network (CNN). In this section, we briefly describe the model architecture and provide explanations of relevant deep-learning and neural network concepts where needed.

To enhance the model performance for each predicted parameter, we trained separate models dedicated to predicting each parameter individually. Nevertheless, to preserve coherence between the predictions, the set of trained models was structured to enable a complete characterization of the SB2 components, attributing each predicted parameter to the corresponding stellar component.

This process was implemented in two stages. First, during the training, the temperature values in the target output array were reordered from lowest to highest. This ensured that the temperature model learned to predict the stellar temperatures consistently. In the second stage, the remaining models were provided with the predicted temperatures (maintaining the same order) and were trained to predict the additional parameters in a manner that was consistent with the established temperature sequence. Consequently, all predicted parameters for each component were consistently presented in the same order.

\subsection{The model architecture}
The networks we used follow the dense CNN architecture inspired by \cite{huang2018densely}, adapted for spectral analysis. Unlike traditional DenseNet applications in image classification \citep{imagenet}, we applied one-dimensional convolutions across the length of the input spectrum. Our models consisted of the following core components: (1) an initial section with a convolutional layer and maximum pooling to extract spectral features, (2) a sequence of dense and transition layers following the DenseNet structure, and (3) fully connected layers for the final prediction.

We trained two separate models: one model to predict the component temperatures from a single SB2 spectrum, and another model to predict the other stellar parameters based on the spectrum and the predicted stellar temperatures. The temperature model relied solely on the extraction of convolutional features, while the model for the other parameters (hereafter, the parameter model) combined spectral features with temperature inputs through fully connected layers. The full architectures are detailed in Table~\ref{tbl:tempmodel} and Table~\ref{tbl:parammodel}.

\begin{table*}[h]
    \centering
    \caption{Step-by-step model architecture in the temperature-prediction model.}
    \begin{tabular}{|c|l|c|c|l|}
        \hline
        \textbf{Step} & \textbf{Layer} & \textbf{Input Shape} & \textbf{Output Shape} & \textbf{Details} \\
        \hline
        1 & Input & $(1, 903)$ & $(1, 903)$ & Raw spectrum \\
        2 & Conv1D & $(1, 903)$ & $(32, 225)$ & Kernel 7, Stride 2, Padding 3 \\
        3 & MaxPool & (32, 450) & (32, 225) & Kernel 3, Stride 2, Padding 1 \\
        4 & Dense Block 1 & $(32, 225)$ & $(128, 225)$ & 2 Conv layers, Kernel 3, BatchNorm, ReLU \\
        4 & Transition Block 1 & $(128, 225)$ & $(64, 112)$ & Conv 1x1, Pool 2x2 \\
        5 & Dense Block 2 & $(64, 112)$ & $(256, 112)$ & 4 Conv layers, Kernel 3, BatchNorm, ReLU \\
        6 & Transition Block 2 & $(256, 112)$ & $(128, 56)$ & Conv 1x1, Pool 2x2 \\
        7 & Dense Block 3 & $(128, 56)$ & $(512, 56)$ & 6 Conv layers, Kernel 3, BatchNorm, ReLU \\
        8 & Transition Block 3 & $(512, 56)$ & $(256, 28)$ & Conv 1x1, Pool 2x2 \\
        9 & Dense Block 4 & $(256, 28)$ & $(512, 28)$ & 4 Conv layers, Kernel 3, BatchNorm, ReLU \\
        10 & Adaptive Pooling & $(512, 28)$ & $(512, 1)$ & Adaptive avg. pooling \\
        11 & Flatten & $(512, 1)$ & $(512)$ & Reshaped for fully connected layers \\
        12 & Fully Connected 1 & $(512)$ & $(512)$ & Linear layer, ReLU \\
        13 & Fully Connected 2 & $(512)$ & $(2)$ & Temperatures, lower first \\
        \hline
    \end{tabular}
    \label{tbl:tempmodel}
\end{table*}

\begin{table*}[h]
    \centering
    \caption{Step-by-step model architecture in the parameter-prediction model (using the spectrum and temperatures as input).}
    \begin{tabular}{|c|l|c|c|l|}
        \hline
        \textbf{Step} & \textbf{Layer} & \textbf{Input Shape} & \textbf{Output Shape} & \textbf{Details} \\
        \hline
        1 & Input (Spectrum) & $(1, 903)$ & $(1, 903)$ & Raw spectrum \\
        2 & Input (Temperatures) & $(2)$ & $(2)$ & Two predicted temperatures (lower first) \\
        3 & Conv1D & $(1, 903)$ & $(32, 450)$ & Kernel 7, Stride 2, Padding 3 \\
        4 & MaxPool & $(32, 450)$ & $(32, 225)$ & Kernel 3, Stride 2, Padding 1 \\
        5 & Dense Block 1 & $(32, 225)$ & $(128, 225)$ & 2 Conv layers, Kernel 3, BatchNorm, ReLU \\
        6 & Transition Block 1 & $(128, 225)$ & $(64, 112)$ & Conv 1x1, Pool 2x2 \\
        7 & Dense Block 2 & $(64, 112)$ & $(256, 112)$ & 4 Conv layers, Kernel 3, BatchNorm, ReLU \\
        8 & Transition Block 2 & $(256, 112)$ & $(128, 56)$ & Conv 1x1, Pool 2x2 \\
        9 & Dense Block 3 & $(128, 56)$ & $(512, 56)$ & 6 Conv layers, Kernel 3, BatchNorm, ReLU \\
        10 & Transition Block 3 & $(512, 56)$ & $(256, 28)$ & Conv 1x1, Pool 2x2 \\
        11 & Dense Block 4 & $(256, 28)$ & $(512, 28)$ & 4 Conv layers, Kernel 3, BatchNorm, ReLU \\
        12 & Adaptive Pooling & $(512, 28)$ & $(512, 1)$ & Adaptive avg. pooling \\
        13 & Flatten & $(512, 1)$ & $(512)$ & Reshaped for fully connected layers \\
        14 & Fully Connected (Temperatures) & $(2)$ & $(64)$ & Linear layer, ReLU \\
        15 & Fully Connected (Merged) & $(512 + 64)$ & $(512)$ & Combined spectral and temperature features \\
        16 & Fully Connected 1 & $(512)$ & $(512)$ & Linear layer, ReLU \\
        17 & Fully Connected 2 & $(512)$ & $(2)$ & Parameters matching sorted temperatures \\
        \hline
    \end{tabular}
    \label{tbl:parammodel}
\end{table*}

\subsection{Loss function}

The objective in training a neural network is to minimize a prescribed loss function that quantifies the difference between the network prediction and the true values. The loss function is evaluated based on the model output, and its error is backpropagated through the network weights. An optimization algorithm \citep[in our case, the Adam optimizer;][]{2014arXiv1412.6980K} adjusts these weights to reduce the loss and improve the predictive accuracy.

We trained two types of neural networks: one network for predicting the temperatures of the two stellar components in a spectrum, and another network for predicting their additional parameters based on the spectrum and the previously inferred temperatures. For both architectures, we used the mean squared error (MSE) loss function, defined as
\begin{equation}
    \text{MSE} = \frac{1}{n} \sum_{i=1}^{n} (y_i - \hat{y}_i)^2,
\end{equation}
where $y_i$ are the true parameter values, and $\hat{y}_i$ are the predicted values. MSE is a standard choice for regression problems because it penalizes larger deviations more heavily than smaller ones. This encourages stable and accurate predictions.

By applying MSE loss separately to each output in both models, we ensured that the network optimized its predictions for the two stellar components independently while preserving the relative ordering constraints (i.e., ensuring that the lower temperature was always predicted first and that predictions for the additional parameters followed the order of the input temperature). For clarity, we present the square root of the MSE when we report and describe the model performance throughout the paper.

During model development, we also experimented with the mean absolute error (MAE) loss function, defined as

\begin{equation}\label{MAE loss}
\centering
    \text{MAE} = \textrm{mean}(|\textrm{error}|) = \frac{1}{n}\sum_{j=1}^n |Y_j - \hat{Y}_j|.
\end{equation}

Since MAE is less sensitive to outliers than other loss functions, it prevents outliers from disproportionately influencing the overall loss. Although we did not use MAE in the final model, we chose to include it in the results section for comparison.

\subsection{Training details}
\label{sub-training_det}
All of our models were trained on the synthetic \textit{Gaia} RVS data, described in Sect.~\ref{sec:sim_data}. The complete synthetic training set consisted of $200$,$000$ spectra. From this dataset, $30$,$000$ spectra were placed in the validation set, and another $20$,$000$ spectra were placed in the test set. The remaining $150$,$000$ spectra were used for the training. Each model was trained for $75$ epochs.

\section{Results}
\label{sec:RES}

\begin{figure*}
\centering
\includegraphics[width=1.7\columnwidth,clip=false]{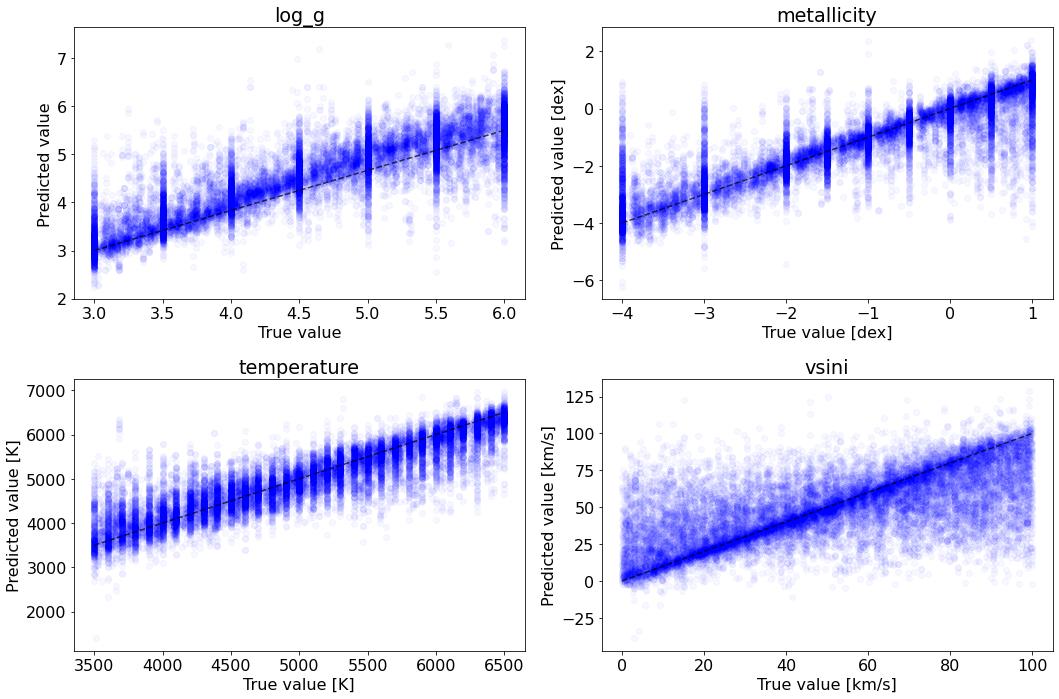}
\caption{Simulated against predicted values in the test set for all predicted parameters.}
\label{fig:sp_graphs}
\end{figure*}

We found that our DenseNet architecture satisfactorily extracted the stellar parameters of SB2 components based on a single spectral observation epoch. The performance for each single parameter model on our $20$,$000$ spectra test set is listed in Table~\ref{table:overall_results} and can also be assessed visually for their prediction-target values in Figure~\ref{fig:sp_graphs}.

\begin{table}
\caption{$\sqrt{\text{MSE}}$ and MAE results for all of our spectral parameter models.}
\label{table:overall_results}
\begin{tabular}{lccc}
\hline
\noalign{\smallskip}
\multicolumn{1}{l}{ } & 
\multicolumn{1}{l}{\text{Range}} &
\multicolumn{1}{l}{$\sqrt{\text{MSE}}$} & 
\multicolumn{1}{l}{MAE} \\
\noalign{\smallskip}
\hline
\noalign{\smallskip}
$T_\text{eff}$               & $3500 - 6500$ [K] & $393.9$ [k] & $283.3$ [K] \\
$\log g$                     & $3.0 - 6.0$   & $0.48$            & $0.32$  \\
$[\element{Fe}/\element{H}]$ & $-4.0 - 1.0$ [dex]   & $0.79$ [dex]            & $0.48$ [dex]  \\
$v \sin i$                   & $0 - 100$ $[\text{km}\,\text{\ s}^{-1}]$      & $22.4$ $[\text{km}\,\text{\ s}^{-1}]$            & $15.4$ $[\text{km}\,\text{\ s}^{-1}]$  \\
% Luminosity                   & $9.8 - 11.7$  & $0.021$            & $0.090$  \\
\noalign{\smallskip}
\hline
\noalign{\smallskip}
\end{tabular}
\end{table}

\begin{table}
\caption{$\sqrt{\text{MSE}}$ and MAE results for the prediction of the difference between the system components.}
\label{table:diff_results}
\begin{tabular}{lccc}
\hline
\noalign{\smallskip}
\multicolumn{1}{l}{ } & 
\multicolumn{1}{l}{\text{Range}} &
\multicolumn{1}{l}{$\sqrt{\text{MSE}}$} & 
\multicolumn{1}{l}{MAE} \\
\noalign{\smallskip}
\hline
\noalign{\smallskip}
$T_\text{eff}$               & $3500 - 6500$ [K] & $678.1$ [K] & $536.2$ [K] \\
$\log g$                     & $3.0 - 6.0$       & $0.79$      & $0.54$      \\
$[\element{Fe}/\element{H}]$ & $-4.0 - 1.0$ [dex] & $1.73$ [dex] & $1.29$ [dex] \\
$v \sin i$                   & $0 - 100$ $[\text{km}\,\text{s}^{-1}]$ & $35.5$ $[\text{km}\,\text{s}^{-1}]$ & $27.3$ $[\text{km}\,\text{s}^{-1}]$ \\
\noalign{\smallskip}
\hline
\noalign{\smallskip}
\end{tabular}
\end{table}

The models were generally successful in assessing the approximate value of the given parameters. Some parameters were more easily learned than others, however. While our model is clearly able to make strong predictions for effective temperature and $\log g$, the model outputs for $v \sin i$ and metallicity exhibited a higher variance. It is important to note that the $v \sin i$ and metallicity prediction MAEs are of a similar scale as the formal precision evaluated by \textit{Gaia} for the rotational velocity \citep{2023A&A...674A...5K} and metallicity \citep{2023A&A...674A..29R} based on the simpler case of single stars, however. More research is needed to evaluate how the accuracy of these predictions can be improved.

In the data presented in Figure~\ref{fig:sp_graphs}, a higher density of points is observed around certain discrete values. This is an expected outcome of the data simulation procedure described in Section~\ref{sec:sim_data}. It results from the grid nature of the PHOENIX data and from the fact that not all parameters were interpolated simultaneously during the data generation in order to preserve the physical consistency of the synthetic spectra.

To further assess the model performance, we examined the prediction of the difference between the two components. This perspective provides an additional insight into evaluating the model behavior and can potentially highlight ordering issues. The results, presented in Table~\ref{table:diff_results}, indicate a satisfactory performance without evidence of these issues. Nevertheless, a comparison with Table~\ref{table:overall_results} shows that our models perform better when each parameter is estimated individually than when the difference between the components is assessed.

\subsection{Impact of the metallicity on the results}

Stellar spectra at low metallicity generally exhibit fewer and weaker absorption lines, which results in a reduced spectral information content. This effect is particularly pronounced at the lower end of our simulated metallicity distribution, where very few identifiable features appear.

\begin{figure}
    \centering
    \includegraphics[width=0.87\columnwidth,clip=false]{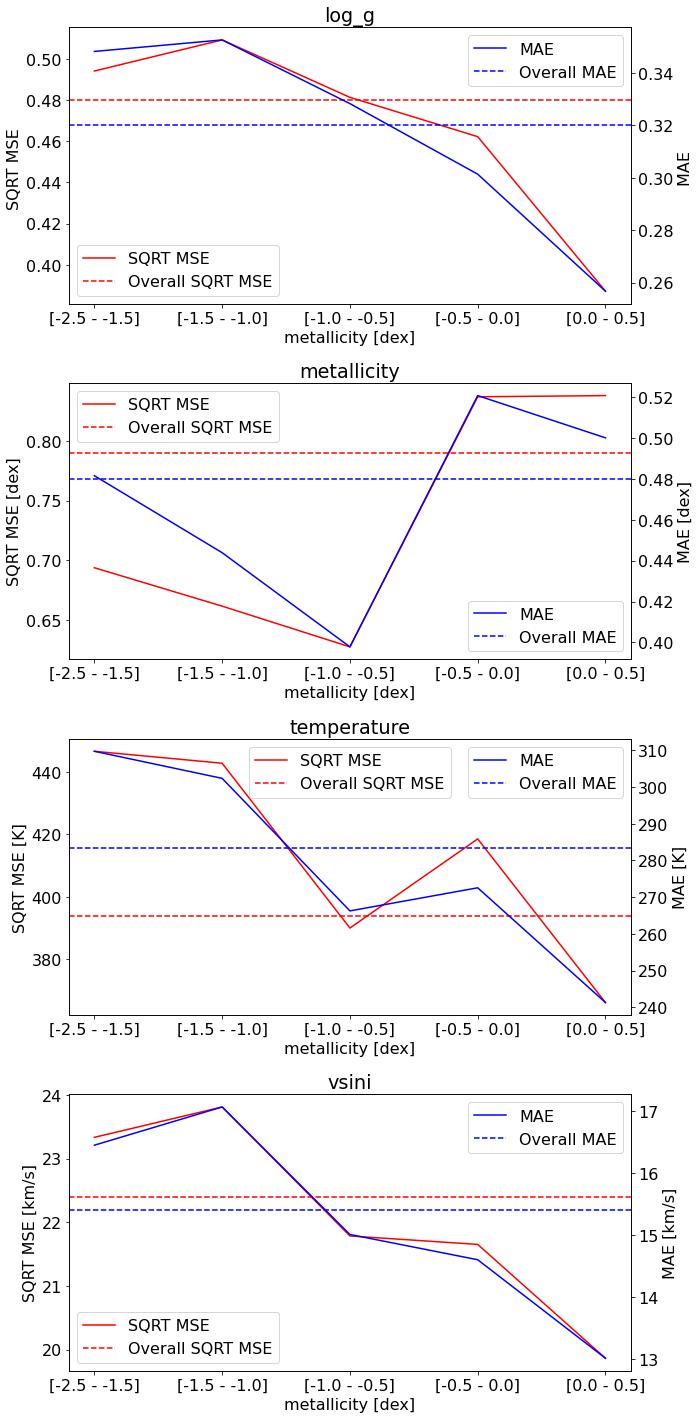}
    \caption{$\sqrt{\text{MSE}}$ and MAE loss vs. metallicity for all stellar parameters.}
    \label{fig:all_met_stuff}
    \end{figure}

To assess the impact of the metallicity on the model performance, we divided the test set into bins based on \([\text{Fe}/\text{H}]\) and evaluated the prediction accuracy within each bin. The performance metrics across the metallicity bins are presented in Figure~\ref{fig:all_met_stuff}. For \(T_\mathrm{eff}\), \(\log g\), and \(v \sin i\), the performance clearly deteriorates with decreasing metallicity. Nonetheless, even in the lowest-metallicity bin, the model accuracy remains reasonable, probably because of a small number of residual features that the model can use.

An exception is the prediction of \([\text{Fe}/\text{H}]\) itself, where the accuracy improves as the metallicity decreases. This may be attributed to the almost complete absence of metal lines, which act as a strong and unambiguous spectral signature for the model to identify. 

\subsection{Impact of the S/N on the results}

Noise significantly affects the inference of the stellar parameters, and higher noise levels obviously complicate the estimation. Our neural networks were trained and tested on data spanning S/N values of $30$–$350$. This enabled us to assess the model performance across this range.

\begin{figure}
    \centering
    \includegraphics[width=0.9\columnwidth,clip=false]{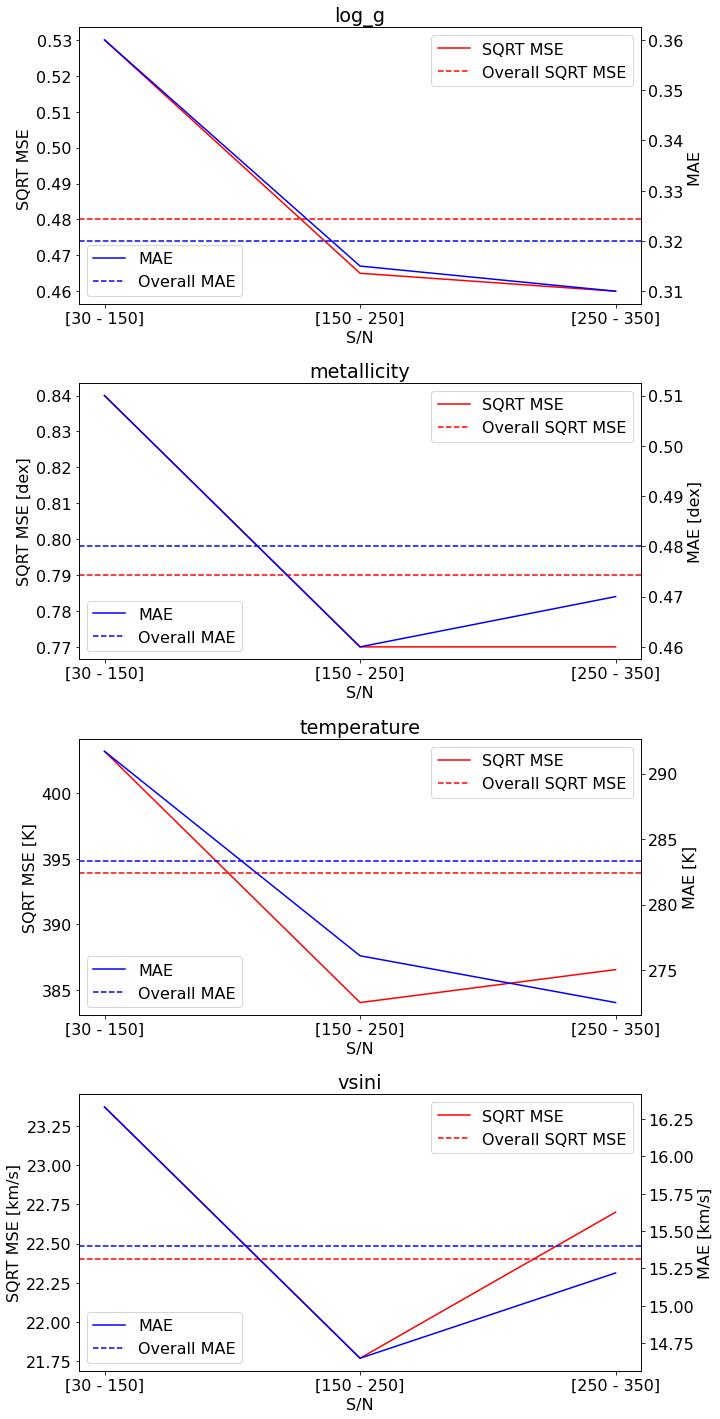}
    \caption{$\sqrt{\text{MSE}}$ and MAE loss vs. S/N for all stellar parameters.}
    \label{fig:all_snr_stuff}
    \end{figure}

As expected, the performance somewhat improved with increasing S/N. Figure~\ref{fig:all_snr_stuff} illustrates this trend and shows that the square root of the MSE and MAE losses decreases with S/N. Notably, the performance improves sharply for an S/N above the $30$–$150$ range, while the highest S/N bucket shows marginally or slightly worse results, which might be due to statistical noise or overfitting.

Most of the real Gaia RVS spectra are expected to exhibit a lower S/N than those we used in the simulated data. While the typical combined spectrum at the end of the mission is expected to reach S/N values of about 20 or slightly higher \citep{GAIADR}, single RVS exposures usually have a much lower S/N, often in the single digits, depending on the magnitude and spectral type of the star \citep{2010EAS....45..189K, 2023A&A...674A...6S}. The actual performance of the RVS instrument should therefore be considered in light of this lower S/N range.

       \begin{figure}
    \centering
    \includegraphics[width=0.9\columnwidth,clip=false]{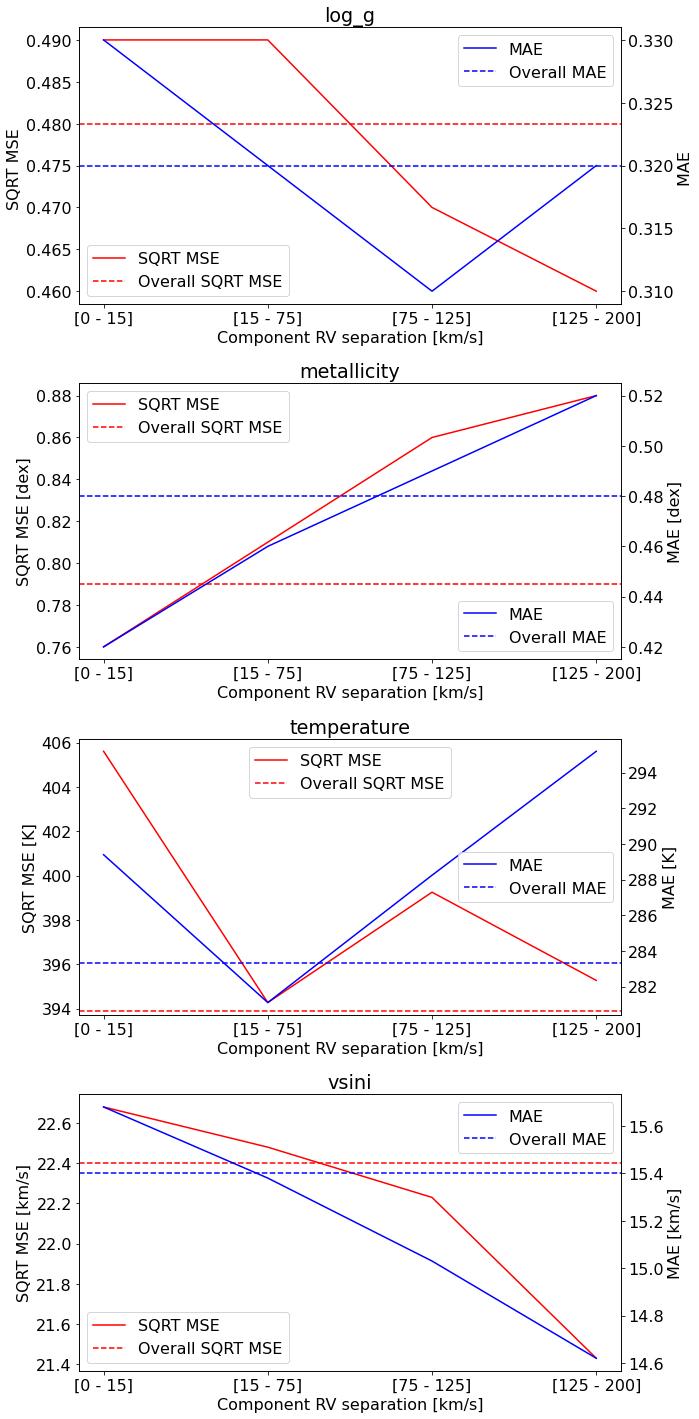}
    \caption{$\sqrt{\text{MSE}}$ and MAE loss vs. component RV separation for all stellar parameters.}
    \label{fig:all_vsep_stuff}
    \end{figure}

\subsection{Impact of the RV separation on the results}
A high RV separation is considered beneficial for the extraction of SB2 parameters because it reduces spectral blending and enhances the feature identification \citep[e.g.,][]{bangies1991, 2019A&A...623A..31S}. Our analysis suggests, however, that the RV separation has little impact on the prediction errors with our method (see Figure~\ref{fig:all_vsep_stuff}). While the errors in $v \sin i$ and $\log g$ slightly decrease with increasing separation, the metallicity errors increase slightly, and the trend for the temperature is mixed. A randomization test using repeated sampling from the test set supported the conclusion that the detected signal is weak and statistically insignificant. Overall, the differences in the RV separation bins are small compared to the typical prediction errors, indicating that it is a secondary factor in our method performance. The reason might be that the well-structured training set fully covered the test set parameter space and minimized the impact of the spectral line separation, in particularly, because the spectral resolution is theoretically sufficient for this task.

We did not attempt to predict the component RVs from the individual spectra because their estimation and interpretation are ambiguous in the context of single observations. We are investigating this approach in a separate project involving multiple spectral observations, however, as outlined in the Discussion section.

\subsection{Results for single-star spectra}
The developed models were specifically designed to analyze SB2 spectra. This raises the question of how they would perform when they are applied to single-star spectra. To assess this, we generated a dedicated test set consisting of single-star spectra and processed it using our models. The results indicate a poor performance: For all parameters, the square root of the MSE and the MAE were approximately twice as high as those obtained for SB2 spectra. Additionally, the inferred parameter sets for the two components exhibited very low correlation coefficients, except for the temperature, where the predicted values matched closely and were highly correlated ($R=\sim0.8$). These findings suggest that our models are poorly suited for single-star spectra, and it highlights the need for tools that work in a preliminary stage \citep[e.g.][]{zhang2021spectroscopic} to identify SB2 spectra before our tool is applied.

\section{Discussion}
\label{sec:diss}

The results we presented demonstrate that deep neural networks can accurately extract the stellar parameters for SB2 components from single-epoch Gaia RVS observations. We propose that this tool be used for the preliminary analysis of SB2 candidates, followed by an analysis of periodic spectral variations to confirm the SB2 classification and refine the parameter extraction.

While it is in principle also possible to include RVs among the predicted parameters, we chose not to do this because RVs can be efficiently and accurately estimated using classical techniques such as a cross-correlation or TODCOR \citep{hill1993, zucker94}. In contrast, the determination of the intrinsic stellar parameters $T_\text{eff}$, $\log g$, $[\element{Fe}/\element{H}]$, and $v \sin i$ is a highly nonlinear inverse problem that depends on subtle, often degenerate features of the spectrum. This makes it particularly well suited for deep neural networks.

The accuracy of the method predictions varies depending on the specific spectral parameter that is estimated, and, naturally, depending on the S/N. Our findings demonstrated that neural networks can accurately predict the temperature and $\log g$, while parameters such as metallicity and $v \sin i$ are more challenging to accurately estimate from a single Gaia RVS spectral observation. In particular, we found that $v \sin i$ consistently shows a lower predictive accuracy than the other parameters. We investigated a range of possible causes, including S/N, luminosity ratio, and metallicity contrast between the components, but found no clear trend. This may reflect an architectural limitation because $v \sin i$ affects the spectral line shapes in a different way than the other parameters. While this should be further explored in future work, the $v \sin i$ estimates remain useful in many contexts, including as initial values for more traditional methods such as TODCOR or for statistical studies. However, given their relatively low accuracy, we do not recommend these estimates for precise or model-sensitive applications, where more reliable methods are necessary.

The prediction of alpha-element abundances was excluded from this study, as discussed in Section \ref{sec:sim_data}. Despite potential challenges that may arise when \textit{Gaia} RVS spectra are used to extract this parameter \citep[see][]{Dafonte_2016}, we anticipate that dedicated future studies that carefully account for this parameter will succeed in predicting it accurately.

Proving to be effective for relatively low-resolution \textit{Gaia} RVS spectra, including in a low S/N regime, our approach can be applied to the analysis of SB2 systems in its RVS spectra \citep{gaia2016}, which will be made available to the community in the coming data releases. This method has several advantages over existing techniques. Primarily, the scanning law of \textit{Gaia} means that some SB2s might have a very limited number of available RVS spectra. In these cases, our proposed technique might be a fast way to identify them as SB2s and characterize their components, without solving for their Keplerian orbits or even extracting their individual RVs.

This highlights another advantage of this approach: its computational efficiency at prediction time. With a trained model, stars can be analyzed in $O(1)$ time, enabling researchers to quickly and efficiently analyze large volumes of spectroscopic data.

One caveat of our newly developed method is that it strongly depends on spectral template models, which are naturally subject to many model assumptions and approximations \citep[see][]{Husetal2013}. It is also interesting to note that certain parameters cannot be physically determined using our approach because of different degeneracies. For example, spectral disentangling alone cannot be used to determine the component flux ratio because a degeneracy exists between the flux ratio and the relative line depths \citep[e.g.,][]{El_Badry_2022}. As mentioned in Sect. \ref{sec:sim_data}, however, the PHOENIX spectral library provides the main-sequence stellar properties derived from the selected parameters, including flux. These properties can therefore be used in this context, for example, to examine the effect of the flux ratio on the accuracy of the results.

Our simulations did not assume identical or even similar metallicities for the two components. Thus, the final predictions allow different metallicities. SB2 cases that would result in significantly different metallicities would require further scrutiny, however, because a result like this would be very surprising and even suspicious, or it might imply quite rare binary formation scenarios \citep[e.g.,][]{2009MNRAS.399.1255M}.

Deep neural networks have not been extensively used in the analysis of \textit{Gaia} RVS spectra so far. Although there are some notable examples, such as \citet{Dafonte_2016}, who demonstrated the use of neural networks for extracting stellar parameters from RVS spectra of single stars, this approach remains relatively underexplored. We hope that this research will raise awareness of the potential and advantages of using deep neural networks for this purpose by highlighting their effectiveness in analyzing complex spectral data.

Other large spectroscopic surveys might also benefit from the application of similar deep-learning-based approaches. These include surveys such as APOGEE \citep{appoggee2017}, LAMOST \citep{lamost12}, the Galactic Archaeology with HERMES survey (GALAH; \citealt{GALAH15}), and the 4-metre Multi-Object Spectroscopic Telescope (4MOST; \citealt{4MOST}). Although these surveys differ significantly from \textit{Gaia} and from one another in terms of their nature and characteristics, they might all gain from an efficient method for extracting the stellar parameters of the component stars, especially for targets with only one or a few exposures.

Similar deep learning methods might also be applied to other aspects of studying SB2 spectra. For instance, we are currently exploring the use of these techniques to directly obtain a full Keplerian solution for SB2 systems from a set of spectral observations. This approach might significantly simplify and accelerate a process that is traditionally complex and computationally intensive.

\begin{acknowledgements}
We are deeply grateful to the referee, Mikhail Kovalev, for his insightful comments, which greatly improved the manuscript.
This research was supported by the ISRAEL SCIENCE FOUNDATION (grant No.\ 1404/22) and the Israel Ministry of Science and Technology (grant No.\ 3-18143). 
The analyses done for this paper made use of the code packages: Astropy \citep{2013A&A...558A..33A, 2018AJ....156..123A, 2022ApJ...935..167A}, NumPy \citep{harris2020array}, SciPy \citep{2020SciPy-NMeth}, PyAstronomy \citep{pya}, and SPARTA \citep{SPARTA2020}.

\end{acknowledgements}

\bibliographystyle{aa}
\bibliography{pPDC}

\end{document}